\documentclass[article]{elsarticle}
\usepackage[utf8]{inputenc}
\usepackage{hyperref}
\usepackage{graphicx}
\usepackage{amsmath}
\usepackage{amssymb,mathrsfs}
\usepackage{lineno}
\usepackage[layout=margin,innerlayout=inline]{fixme}
\usepackage{tikz}
\usepackage{caption}
\usepackage{subcaption}
\usepackage{hyperref}
\usepackage{makecell}
\usepackage{multirow}
\usepackage{longtable}
\usepackage{fancyhdr}
\usepackage{array}
\usepackage{makecell}
\usepackage{slashbox}
\usepackage{booktabs}

\newcolumntype{C}[1]{>{\centering\arraybackslash}p{#1}}

\makeatletter
\def\ps@pprintTitle{%
 \let\@oddhead\@empty
 \let\@evenhead\@empty
 \def\@oddfoot{\centerline{\thepage}}%
 \let\@evenfoot\@oddfoot}
\makeatother
\modulolinenumbers[10]

\journal{Preprint}

\begin{document}

\begin{frontmatter}
\title{The reliability of a deep learning model in clinical out-of-distribution MRI data: a multicohort study}

\author[ki]{Gustav Mårtensson\corref{mycorrespondingauthor}}%
\ead{gustav.martensson@ki.se}
\author[ki]{Daniel Ferreira }
\author[ki2,ks]{Tobias Granberg }
\author[ki2,ks]{Lena Cavallin }
\author[nor1,nor2,nor3]{Ketil Oppedal }
\author[bre]{Alessandro Padovani }
\author[brno]{Irena Rektorova }
\author[chi]{Laura Bonanni }
\author[gen]{Matteo Pardini }
\author[lju]{Milica Kramberger }
\author[new]{John-Paul Taylor }
\author[pra]{Jakub Hort }
\author[rek]{J\'{o}n Sn{\ae}dal }
\author[stp1,stp2,stp3,stp4]{Jaime Kulisevsky }
\author[str1,str2]{Frederic Blanc }
\author[ven]{Angelo Antonini }

\author[a2]{Patrizia Mecocci}
\author[a3]{Bruno Vellas}
\author[a4]{Magda Tsolaki}
\author[a5]{Iwona K\l{}oszewska}
\author[a6,a7]{Hilkka Soininen}
\author[a8]{Simon Lovestone}
\author[a9,a10,london3]{Andrew Simmons}
\author[nor1,london4]{Dag Aarsland}
\author[ki,london3]{Eric Westman}
\author[]{for the Alzheimer’s Disease Neuroimaging Initiative \tnoteref{adni}}

\address[ki]{\scriptsize Division of Clinical Geriatrics, Department of Neurobiology, Care Sciences and Society, Karolinska Institutet, Stockholm, Sweden.}
\address[ki2]{Department of Clinical Neuroscience, Karolinska Institutet, Stockholm, Sweden.}
\address[ks]{Department of Radiology, Karolinska University Hospital, Stockholm, Sweden.}

\address[nor1]{Centre for Age-Related Medicine, Stavanger University Hospital, Stavanger, Norway}
\address[nor2]{Department of Radiology, Stavanger University Hospital, Stavanger, Norway}
\address[nor3]{Department of Electrical Engineering and Computer Science, University of Stavanger, Stavanger, Norway}
\address[bre]{Neurology Unit, Department of Clinical and Experimental Sciences, University of Brescia, Brescia, Italy}
\address[brno]{1st Department of Neurology, Medical Faculty, St. Anne’s Hospital and CEITEC, Masaryk University, Brno, Czech Republic}
\address[chi]{Department of Neuroscience Imaging and Clinical Sciences and CESI, University G d’Annunzio of Chieti-Pescara, Chieti, Italy}
\address[gen]{Department of Neuroscience (DINOGMI), University of Genoa and Neurology Clinics, Polyclinic San Martino Hospital, Genoa, Italy}
\address[lju]{Department of Neurology, University Medical Centre Ljubljana, Medical faculty, University of Ljubljana, Slovenia}
\address[new]{Institute of Neuroscience, Newcastle University, Newcastle upon Tyne, UK}
\address[pra]{Memory Clinic, Department of Neurology, Charles University, 2nd Faculty of Medicine and Motol University Hospital, Prague, Czech Republic}
\address[rek]{Landspitali University Hospital, Reykjavik, Iceland}
\address[stp1]{Movement Disorders Unit, Neurology Department, Sant Pau Hospital, Barcelona, Spain}
\address[stp2]{Institut d'Investigacions Biom\'ediques Sant Pau (IIB-Sant Pau), Barcelona, Spain}
\address[stp3]{Centro de Investigaci\'on en Red-Enfermedades Neurodegenerativas (CIBERNED), Barcelona, Spain}
\address[stp4]{Universitat Aut\'onoma de Barcelona (U.A.B.), Barcelona, Spain}
\address[str1]{Day Hospital of Geriatrics, Memory Resource and Research Centre (CM2R) of Strasbourg, Department of Geriatrics, H\^opitaux Universitaires de Strasbourg, Strasbourg, France}
\address[str2]{University of Strasbourg and French National Centre for Scientific Research (CNRS), ICube Laboratory and F\'ed\'eration de M\'edecine Translationnelle de Strasbourg (FMTS), Team Imagerie Multimodale Int\'egrative en Sant\'e (IMIS)/ICONE, Strasbourg, France}
\address[ven]{Department of Neuroscience, University of Padua, Padua \& Fondazione Ospedale San Camillo, Venezia, Venice, Italy}

\address[a2]{Institute of Gerontology and Geriatrics, University of Perugia, Perugia, Italy.}
\address[a3]{UMR INSERM 1027, gerontopole, CHU, University of Toulouse, France.}
\address[a4]{3rd Department of Neurology, Memory and Dementia Unit, Aristotle University of Thessaloniki, Thessaloniki, Greece.}
\address[a5]{Medical University of Lodz, Lodz, Poland.}
\address[a6]{Institute of Clinical Medicine, Neurology, University of Eastern Finland.}
\address[a7]{Neurocenter, Neurology, Kuopio University Hospital, Kuopio, Finland.}
\address[a8]{Department of Psychiatry, Warneford Hospital, University of Oxford, Oxford, UK.}
\address[a9]{NIHR Biomedical Research Centre for Mental Health, London, UK.}
\address[a10]{NIHR Biomedical Research Unit for Dementia, London, UK.}
\address[london3]{Department of Neuroimaging, Centre for Neuroimaging Sciences, Institute of Psychiatry, Psychology and Neuroscience, King’s College London, London, UK.}
\address[london4]{Institute of Psychiatry, Psychology and Neuroscience, King’s College London, London, UK}
\cortext[mycorrespondingauthor]{Corresponding author}

\tnotetext[adni]{\tiny Data used in preparation of this article were obtained from the Alzheimer’s Disease Neuroimaging Initiative (ADNI) database (adni.loni.usc.edu). As such, the investigators within the ADNI contributed to the design and implementation of ADNI and/or provided data but did not participate in analysis or writing of this report. A complete listing of ADNI investigators can be found at: \url{http://adni.loni.usc.edu/wp-content/uploads/how_to_apply/ADNI_Acknowledgement_List.pdf}.}

\begin{abstract}
Deep learning (DL) methods have in recent years yielded impressive results in medical imaging, with the potential to function as clinical aid to radiologists. However, DL models in medical imaging are often trained on public research cohorts with images acquired with a single scanner or with strict protocol harmonization, which is not representative of a clinical setting. The aim of this study was to investigate how well a DL model performs in unseen clinical data sets---collected with different scanners, protocols and disease populations---and whether more heterogeneous training data improves generalization. In total, 3117 MRI scans of brains from multiple dementia research cohorts and memory clinics, that had been visually rated by a neuroradiologist according to Scheltens' scale of medial temporal atrophy (MTA), were included in this study. By training multiple versions of a convolutional neural network on different subsets of this data to predict MTA ratings, we assessed the impact of including images from a wider distribution during training had on performance in external memory clinic data. Our results showed that our model generalized well to data sets acquired with similar protocols as the training data, but substantially worse in clinical cohorts with visibly different tissue contrasts in the images. This implies that future DL studies investigating performance in out-of-distribution (OOD) MRI data need to assess multiple external cohorts for reliable results. Further, by including data from a wider range of scanners and protocols the performance improved in OOD data, which suggests that more heterogeneous training data makes the model generalize better. To conclude, this is the most comprehensive study to date investigating the domain shift in deep learning on MRI data, and we advocate rigorous evaluation of DL models on clinical data prior to being certified for deployment.

\end{abstract}

\end{frontmatter}

%\linenumbers

\section{Introduction}
\begin{figure}[h!]
\centerline{
\includegraphics[width=1\textwidth]{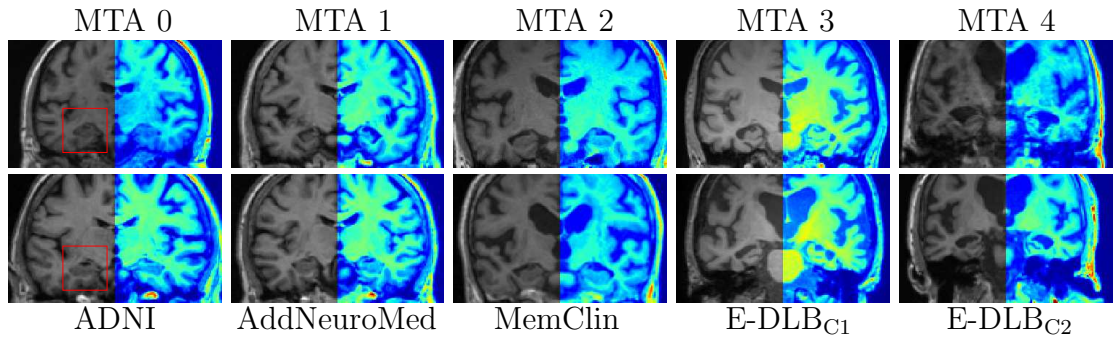}
}
\caption{\label{fig:cohort_ex} Two randomly selected images from five different cohorts in the study to illustrate image intensity variability between cohorts, and examples of Scheltens' scale of medial temporal atrophy (MTA) rated by a radiologist. The red boxes show the region of interest for the MTA scale with a progressive worsening in the hippocampus and surrounding regions. The images are normalized to have zero mean and unit variance, with the same intensity color scale for all images. The jet color map on the right-hand part of the images is used to visibly highlight intensity differences between centers.}
\end{figure}

The use of deep learning (DL) models in neuroimaging has increased rapidly in the last few years, often showing superior diagnostic abilities compared to traditional imaging softwares (see \cite{Litjens2017,Lundervold2019} for reviews). This makes DL models promising to use as diagnostic aid to clinicians. However, for a software to function in a clinical setting it should work on images acquired from different scanners, protocol parameters, and of varying image quality---a scenario reflective of most clinical settings today. Fig. \ref{fig:cohort_ex} shows illustrative examples of the variability in images from some different centers included in this study.

Training a DL model on magnetic resonance imaging (MRI) scans requires a large dataset to obtain good performance. However, (labeled) clinical data is generally difficult (and expensive) to acquire due to strict privacy regulations on medical data. Most researchers are therefore constrained to rely on publicly available neuroimaging datasets, which are typically research cohorts that differ from a clinical setting in several ways: 1) Images are acquired from the same scanner and protocol, or protocols have been harmonized across machines. This is done to reduce image variability and confounding effects, which are problematic also for traditional neuroimaging softwares such as FSL, FreeSurfer and SPM \cite{Guo2019}. 2) Research cohorts often have strict inclusion and exclusion criteria for the individuals enrolled in order to study a particular effect of interest. For instance, to study the progression of patients suffering from Alzheimer's Disease (AD) it may be necessary to exclude comorbidities, such as cerebrovascular pathology or history traumatic brain injury, in order to reduce heterogeneity not relevant to the research question. This is the case of the Alzheimer's Disease Neuroimaging Initiative (ADNI) cohort---the most extensive public neuroimaging data set in AD and used for training and evaluation in multiple DL studies on AD \cite{Litjens2017}. However, since comorbidities are frequent alongside AD the ADNI cohort is hardly reflective of the heterogeneous AD profiles of patients in the clinics \cite{Boyle2018,Ferreira2017a}. Thus, training a DL model on data from a research cohort may perform worse in a clinical setting due to difficulties generalizing to new scanners/protocols (point 1) and/or more heterogeneous population (point 2). Investigating the performance in \textit{out-of-distribution data} (OOD data, i.e. images acquired with different scanners/protocols than the ones included in the training set) is an important step in order to investigate clinical applicability of DL models and understanding the challenges that can arise when deploying.

Some previous studies have investigated the clinical applicability of machine learning models, or \textit{domain-shift} (training a model on data from one domain and applying it in data from another). A recent paper by De Fauw et al. (2018) trained and applied a deep learning model on a clinical dataset of 3D optical coherence tomography scans, which managed to predict referral decisions with similar performance as experts \cite{DeFauw2018}. However, when applied to images from a new scanning device the performance was poor. Since they used a two-stage model architecture, where the first part segmented the image into different tissue types (making subsequent analysis scanner independent), it was sufficient to retrain only the segmentation network with a (much smaller) data set from the new device. Klöppel and colleagues (2015) investigated the performance of a trained SVM-classifier to diagnose dementia in a clinical data set of a more heterogeneous population \cite{Kloppel2015}. Their models were also fed tissue-segmentation maps, preprocessed using SPM, and found a drop in performance compared to the "clean" training set, as well as lower performance than previous studies had reported (typically cross-validation performance). Zech et al. (2018) explicitly investigated how a convolutional neural network (CNN) trained for pneumonia screening on chest X-rays generalized to new cohorts. They found significantly lower performance in OOD cohorts. Further, they demonstrated that a CNN could accurately classify which hospital an image was acquired at and thus potentially leverage this information to adjust the prediction method due to different disease prevalences in the cohorts \cite{Zech2018}. Some recent studies have investigated MRI segmentation performance across centers and again found drops in performance \cite{Kamnitsas2017,Perone2019,Albadawy2018}. These analyses were made on a small number of images, as segmented data is typically expensive and time-consuming to label. In contrast to segmented data, visual ratings of atrophy, which still serve as the main tools to quantify neurodegeneration in memory clinics, offer a faster method to annotate brain images that make it feasible to label large datasets ($>$1000 images) from multiple clinics. Our group recently proposed \textit{AVRA} (Automatic Visual Ratings of Atrophy), a DL model based on convolutional neural networks (CNN) \cite{Martensson2018b}. AVRA inputs an unprocessed T$_1$-weighted MRI image and predicts the ratings of Scheltens' Medial Temporal Atrophy (MTA) scale, commonly used clinically to diagnose dementia \cite{Scheltens1992} (see Fig. \ref{fig:cohort_ex} for examples of the MTA scale).

The aim of this study is to systematically investigate the performance of a CNN based model (AVRA) in OOD data from clinical neuroimaging cohorts. We study the impact more heterogeneous training data has on generalization to OOD data by training and evaluating AVRA on images from different combinations of cohorts. Two of these cohorts are research oriented: similar to each other in terms of disease population (AD) and protocol harmonization. The other two datasets consist of clinical data from multiple European sites including individuals of different and mixed types of dementia, not just AD. Additionally, we assess the inter- and intra-scanner variability of AVRA in a systematic test-retest set. To our knowledge this is the largest and most comprehensive study on the generalization of DL models in neuroimaging and MRI data. 

\section{Material and methods}
\subsection{MRI data and protocols}
The 3117 images analyzed in this study came from five different cohorts described in Table \ref{tab:cohorts}, where we also list the reasons for including these datasets in the current study. Full lists of scanners and scanning protocols are provided as Supplementary Data. TheHiveDB was used for data management in this study \cite{Muehlboeck2014}.

\begin{table}[h!]
\caption{\label{tab:cohorts} An overview of how the cohorts used for training and/or evaluation differ from each other, and the purpose of including them in the present study. The E-DLB cohort (denoted as E-DLB$_{\text{all}}$, referring to all images in the cohort) was stratified into different subsets in order to isolate specific features of interest. N$_{\text{train}}$/N$_{\text{test}}$ refers to the number of labeled images used during training/evaluation, where some cohorts were split into training and test set. Abbreviations: Deep Learning (DL); Out-of-distribution (OOD) data; Alzheimer's disease (AD); Healthy controls (HC); Frontotemporal lobe dementia (FTLD); Dementia with Lewy Bodies (DLB); Parkinson's disease with dementia (PDD).}

\small
\begin{tabular}{m{2.3cm} m{3.1cm} m{3.3cm} m{4.1cm}}
\toprule
\textbf{Cohort} & \textbf{Scanners/Protocols} & \textbf{Disease population} & \textbf{Purpose of inclusion} \\ \midrule
ADNI \hspace{1cm} \footnotesize{N$_{\text{train}}$=1568 N$_{\text{test}}$=398} & Multiple scanners and sites, but strictly harmonized with phantom. Both 1.5T and 3T. & AD spectrum and HC. & Common cohort to train and evaluate DL models in, which we hypothesize should not generalize well.\\ 
\addlinespace
AddNeuroMed \footnotesize{N=122} & Harmonized, designed to be compatible with ADNI. & AD patients only. & Assess AVRA in an external research cohort similar to ADNI. \\ 
\addlinespace
MemClin \footnotesize{N$_{\text{train}}$=318 N$_{\text{test}}$=66} & Unharmonized, part of clinical routine from a single memory clinic. & Mainly AD spectrum and HC, with 37 FTLD patients. & Large clinical cohort with similar disease population as ADNI and AddNeuroMed. \\ 
\addlinespace
E-DLB$_{\text{all}}$ \hspace{1cm}\footnotesize{N=645} & Retrospective unharmonized data of varying quality from 12 European sites as part of their clinical routine. & Mainly DLB spectrum, but also HC, AD and PDD. & To assess performance of AVRA in a large, realistic clinical cohort. \\ 
\addlinespace
E-DLB$_{\text{AD}}$ \footnotesize{N=193} & Same as E-DLB$_{\text{all}}$ & Only individuals with AD pathology from E-DLB$_{\text{all}}$. & To isolate effects of scanners/protocols not seen during training from disease population. \\ 
\addlinespace
E-DLB$_{\text{DLB,PDD}}$ \footnotesize{N=\{266,97\}} & Same as E-DLB$_{\text{all}}$ & Only individuals with DLB or PDD pathology from E-DLB$_{\text{all}}$, respectively. & To assess the impact scanners/protocols \emph{and} disease populations not seen during training have on AVRA performance. \\ 
\addlinespace
E-DLB$_{\text{25,50\%}}$ \footnotesize{N$_{\text{train}}$=\{173,312\} N$_{\text{test}}$= 333} & Same as E-DLB$_{\text{all}}$ &  Randomly selected images with a probability of 25\% (or 50\%) from all centers in E-DLB$_{\text{all}}$. & To assess effect of including training data from test set distribution has on AVRA performance. \\  
\addlinespace
E-DLB$_{\text{C1,C2}}$ \footnotesize{N$_{\text{train}}$=379 N$_{\text{test}}$=\{101,165\}} & Both centers have used a single scanner (3T) and protocol. & Stratifying images into three groups: from center C1, from C2, and all images \emph{not} in C1, C2 from E-DLB$_{\text{all}}$. & "External validation sets": how would AVRA perform if deployed in two external memory clinics?\\
\addlinespace
Test-Retest \footnotesize{N=72} & Three Siemens scanners (two 1.5T, one 3T) with similar protocols but unharmonized. & Young (38 $\pm$ 13 years old) MS patients and healthy controls. & Systematic evaluation of the impact scanner variability has on AVRA predictions. \\
\bottomrule
\end{tabular}
\end{table}

Data used in the preparation of this article were obtained from the Alzheimer’s Disease Neuroimaging Initiative (ADNI) database (\url{adni.loni.usc.edu}). The ADNI was launched in 2003 as a public-private partnership, led by Principal Investigator Michael W. Weiner, MD. The primary goal of ADNI has been to test whether serial magnetic resonance imaging (MRI), positron emission tomography (PET), other biological markers, and clinical and neuropsychological assessment can be combined to measure the progression of mild cognitive impairment (MCI) and early Alzheimer’s disease (AD). In brief, the ADNI dataset is a large, public dataset that has been helped advance the field of AD and neuroimaging. However, the strictly harmonized protocols and strict exclusion criteria make ADNI unrepresentative of a clinical setting. Some subjects were scanned multiple times (within a month) in both a 1.5T and a 3T scanner in which case one of the images was selected at random during training for the current study. \textit{AddNeuroMed} is an imaging cohort collected in six sites across Europe with the aim of developing and validating novel biomarkers for AD \cite{Simmons2011}. The MRI images were acquired with protocols designed to be compatible with data from ADNI, and the two cohorts have been successfully combined in previous studies \cite{Westman2011a,Falahati2016,Martensson2018}. AddNeuroMed was an interesting cohort to assess AVRA's reliability in due to having consistent scanning parameters and acquisition methods similar to ADNI. Thus, this dataset represented a research cohort where we expected our DL model to show good performance in when trained on ADNI data. A subset of the images (122) of patients diagnosed with AD had been visually rated for MTA. Exclusion criteria for both these studies included no histories of head trauma, neurological or psychiatric disorders (apart from AD), organ failure, or drug/alcohol abuse. 

The \textit{MemClin} data set was used for training also in our previous study detailing AVRA \cite{Martensson2018b}. MemClin consists of images of AD or frontotemporal lobe dementia collected from the memory clinic at Karolinska Hospital in Huddinge, Sweden. This data set better resembled a clinical setting with varying scanning parameters and field strengths, while the disease population was not completely representative of patients in a memory clinic. The only exclusion criteria was history of traumatic brain injury. Images and ratings have previously been analyzed in \cite{Ferreira2018,Lindberg2009}.

The fourth cohort in this study consists of clinical MRI images from the European consortium for Dementia with Lewy Bodies (referred to as \textit{E-DLB} from here on) previously described in \cite{Kramberger2017,Oppedal2019}. Patients with referrals to memory, movement disorders, psychiatric, geriatric or neurology clinics that had undergone an MRI were selected from 12 sites in Europe. These individuals were diagnosed with Dementia with Lewy Bodies (DLB), AD, Parkinson's Disease with Dementia (PDD), mild cognitive impairment (MCI, due to AD or DLB), or were normal elderly controls (NC). The images were acquired as part of the clinical routine, and consequently without protocol harmonization, and can thus be considered to reflect a clinical setting well. Exclusion criteria for the E-DLB cohort were having received a recent diagnosis of major somatic illness, history of psychotic or bipolar disorders, acute delirium, or terminal illness.

We also investigated AVRA's rating consistency on unprocessed MRI images (i.e. no lesion filling) of three healthy and nine individuals with Multiple Sclerosis (MS, mean disease duration 7.3$ \pm$ 5.2 years) that were scanned twice with repositioning in three different Siemens scanners (i.e. six scans in total) in a single day. Six of the patients had relapsing-remitting MS, two secondary progressive MS, and one primary progressive MS. This data set was collected for a previous study \cite{Guo2019}, and we will refer to this small set as the \textit{test-retest} dataset. These individuals were not rated for MTA by a radiologist.

\subsection{Radiologist ratings}
An experienced neuroradiologist (Lena Cavallin, L.C.) visually rated 3117 T$_1$-weighted brain images (blind to age and sex) according to the established MTA rating scale. These ratings have been used in previous studies on AD \cite{Ferreira2015} and E-DLB \cite{Oppedal2019} by our group, and the distribution of ratings are shown in Table \ref{tab:dist}. These rating scales provide a quantitative measure of atrophy in specific regions, and while they are often used for dementia diagnosis the rating scales themselves are independent of diagnosis, age and sex. L.C. has previously demonstrated excellent inter- and intra-rater agreements in research studies \cite{Martensson2018b}.

\begin{table}[h!]
\caption{\label{tab:dist} Distribution of MTA ratings from a neuroradiologist in the different cohorts, together with sex (female percentage) and age demographics. The lines in bold refers to the statistics of the \emph{whole} cohort, whereas the rows not in boldface text are the subsets used for during training. $N$ is the total number of rated images, and since both left and right hemispheres were rated there were $2N$ ratings. \textit{MTA distribution} shows the percentage of each radiologist rating per (sub-)cohort. A small linespace are added between some E-DLB subsets to illustrate the grouping of the subsets where no overlap between training and test sets occur. }
\centerline{
\footnotesize
\begin{tabular}{*{7}l c c}
\toprule
\multicolumn{1}{l}{\textbf{Cohort}} & \multicolumn{1}{c}{N} & \multicolumn{5}{c}{MTA distribution, (\%)} & \multicolumn{1}{c}{Females} & \multicolumn{1}{c}{Age} \\ 
\cmidrule(lr){3-7}
\multicolumn{1}{l}{\hspace{.3cm} Subset} &  & \multicolumn{1}{c}{0} & \multicolumn{1}{c}{1} & \multicolumn{1}{c}{2} & \multicolumn{1}{c}{3} & \multicolumn{1}{c}{4} & \multicolumn{1}{c}{(\%)} & \multicolumn{1}{c}{(mean $\pm$ std)} \\ 
\midrule 
\multicolumn{1}{l}{\textbf{ADNI$_\text{all}$}} & 1966  & 11 & 40 & 29 & 14 & 6 & 41 & 76.9 $\pm$ 6.6\\ 
\multicolumn{1}{l}{\hspace{.3cm} ADNI$_\text{train}$} & 1568  & 11 & 40 & 29 & 14 & 6 & 41 & 77.0 $\pm$ 6.6\\ 
\multicolumn{1}{l}{\hspace{.3cm} ADNI$_\text{test}$} & 398  & 12 & 39 & 28 & 16 & 5 & 43 & 76.6 $\pm$ 6.9\\ 
\addlinespace
\multicolumn{1}{l}{\textbf{AddNeuroMed}} & 122  & 2 & 21 & 41 & 23 & 13 & 66 & 75.7 $\pm$ 6.1\\ 
\addlinespace
\multicolumn{1}{l}{\textbf{MemClin$_\text{all}$}} & 384  & 3 & 35 & 39 & 18 & 6 & 57 & 68.0 $\pm$ 8.2\\ 
\multicolumn{1}{l}{\hspace{.3cm} MemClin$_\text{train}$} & 318  & 3 & 34 & 40 & 17 & 6 & 56 & 68.0 $\pm$ 8.2\\ 
\multicolumn{1}{l}{\hspace{.3cm} MemClin$_\text{test}$} & 66  & 4 & 39 & 33 & 21 & 4 & 61 & 68.3 $\pm$ 8.2\\ 
\addlinespace
\multicolumn{1}{l}{\textbf{E-DLB$_\text{all}$}} & 645  & 14 & 41 & 29 & 12 & 4 & 44 & 73.7 $\pm$ 8.0\\ 
\multicolumn{1}{l}{\hspace{.3cm} E-DLB$_\text{25}^{\text{train}}$} & 149  & 15 & 40 & 28 & 12 & 4 & 43 & 74.2 $\pm$ 8.1\\ 
\multicolumn{1}{l}{\hspace{.3cm} E-DLB$_\text{50}^{\text{train}}$} & 324  & 15 & 41 & 29 & 11 & 3 & 45 & 74.0 $\pm$ 8.1\\ 
\multicolumn{1}{l}{\hspace{.3cm} E-DLB$_\text{50}^{\text{test}}$} & 321  & 12 & 42 & 29 & 12 & 5 & 43 & 73.4 $\pm$ 8.0\\ 
\addlinespace
\multicolumn{1}{l}{\hspace{.3cm} E-DLB$_\text{C1,C2}^{\text{train}}$} & 379  & 11 & 42 & 30 & 15 & 3 & 51 & 73.7 $\pm$ 7.5\\ 
\multicolumn{1}{l}{\hspace{.3cm} E-DLB$_\text{C1}$} & 101  & 16 & 40 & 29 & 11 & 4 & 23 & 75.9 $\pm$ 6.5\\ 
\multicolumn{1}{l}{\hspace{.3cm} E-DLB$_\text{C2}$} & 165  & 19 & 41 & 28 & 6 & 5 & 41 & 72.3 $\pm$ 9.5\\ 
\addlinespace	
\multicolumn{1}{l}{\hspace{.3cm} E-DLB$_\text{AD}$} & 193  & 4 & 30 & 38 & 20 & 7 & 55 & 75.7 $\pm$ 7.7\\ 
\multicolumn{1}{l}{\hspace{.3cm} E-DLB$_\text{DLB}$} & 266  & 14 & 43 & 28 & 11 & 4 & 44 & 73.6 $\pm$ 8.2\\ 
\multicolumn{1}{l}{\hspace{.3cm} E-DLB$_\text{PDD}$} & 97  & 19 & 46 & 27 & 7 & 1 & 15 & 71.8 $\pm$ 7.0\\ 
\bottomrule
\end{tabular}}
\end{table}

\subsection{Model description}
Our group recently proposed a method we call \textit{AVRA} (Automatic Visual Ratings of Atrophy) that provides computed scores of three visual rating scales commonly used clinically: Scheltens' MTA scale (see Fig. \ref{fig:cohort_ex}), Pasquier's frontal subscale of global cortical atrophy (GCA-F), and Koedam's scale of posterior atrophy (PA) \cite{Martensson2018b}. AVRA showed substantial rating-agreement to an expert neuroradiologist in all three scales on a hold-out test set ($N$=464) that was drawn from the same distribution as the training data ($N$=1886) from two AD cohorts. Since the measures are independent of diagnosis, sex and age, a DL tool such as AVRA (trained end-to-end and does its own feature-extraction from the entire brain volume) should work equally well on different disease populations.

For this experiment we focused only on the MTA scale and used the same network architecture and hyperparameters as previously described in \cite{Martensson2018b}, but with different combinations of cohorts in the training set. Briefly, AVRA is a Recurrent Convolutional Neural Network (R-CNN) that inputs an unprocessed MRI volume, which is then processed slice-by-slice by the model. A residual attention network \cite{Wang2017} is used to extract features from each slice, and these are forwarded to a Long-Short Term Memory (LSTM) network \cite{Hochreiter1997}. The LSTM modules remember relevant information provided from each slice and use it to predict the atrophy score the radiologist would give. This prediction is continuous, but when studying the inter-rater agreement with the radiologist, expressed in kappa statistics or accuracy, we round AVRA's output to the nearest integer.

A trained version of AVRA is publicly available targeted towards neuroimaging researchers at \url{https://github.com/gsmartensson/avra_public}.

\subsection{Training procedure}
To systematically investigate the performance in new data distributions we trained versions of AVRA on data where we kept the number of subjects fixed to the maximum size of the ADNI training data ($N$ = 1568), since more training data generally leads to better performance and could bias the results. ADNI was the largest dataset with ratings available to us, and needed to be part of all training sets in order for the number of images to be large enough for training. When combining data from an additional cohort, we replaced a subject in ADNI with one from the new cohort that had the same ratings from the radiologist. This way, both the size and the distribution of the training data were kept constant. Each training set was divided into five cross-validation sets (to replicate the procedure in \cite{Martensson2018b}) and these five trained models were used as an ensemble classifier. 

Each of the cohorts have different characteristics, as outlined in Table \ref{tab:cohorts}. Since the E-DLB cohort was highly diverse in terms of scanners and disease population, we stratified it into different partitions (some with overlap, but no training/test set pairs shared any images) in order to isolate specific features. To investigate the performance drop due to OOD test data, we randomly assigned each subject into E-DLB$_{\text{25\%}}^{\text{train}}$, E-DLB$_{\text{50\%}}^{\text{train}}$ and E-DLB$_{\text{50\%}}^{\text{test}}$, where the numbers refer to the percentage of subject from the whole cohort and with no overlap between $^{\text{train}}$ and $^{\text{test}}$. This setup aims to simulate realistic ways of introducing a DL model into a new clinic: 1) \textit{as is} (i.e. no additional labeled data from the new clinic), 2) retraining, or finetuning, the existing model with \textit{some} additional labeled data from the same clinics (E-DLB$_{\text{25\%}}$), 3) same as 2) but with twice as much additional data (E-DLB$_{\text{50\%}}^{\text{train}}$).

To assess the impact of disease population we sampled individuals on the AD spectrum (E-DLB$_{\text{AD}}$), DLB spectrum (E-DLB$_{\text{DLB}}$), or with PDD (E-DLB$_{\text{PDD}}$) into three subsets. Since the main bulk of training images comes from ADNI---an AD cohort---it is of interest to see if the models overfit to AD atrophy patterns and are influenced by neighboring regions in the medial temporal lobe not part of the MTA scale. 

To study if AVRA's generalizability improved when widening the training data distribution we also computed the performance on data from two clinics that we refer to as E-DLB$_{\text{C1}}$ and E-DLB$_{\text{C2}}$. A single 3T scanner and protocol was used at each site for scanning, yet with visibly different image intensities (see image examples in Fig. \ref{fig:cohort_ex}). We view these centers as "external validation sets" to estimate the performance we may expect if implementing AVRA in a new memory clinic (although single-scanner usage and study populations may not perfectly represent a memory clinic sample). We included data from all other centers to our training set (E-DLB$_{\text{C1,C2}}^{\text{train}}$) to study if more heterogeneous training data improves generalization to new protocols. 

\subsection{Evaluation metrics}
We assess the performance of AVRA using Cohen's linearly weighted kappa $\kappa_w$, which is the most common metric to assess inter- and intra-rater agreement for visual ratings in the literature. It ranges from [-1,1] where $\kappa_w$ $\in$ [0.2,0.4) is generally considered \textit{fair}, $\kappa_w$ $\in$ [0.4,0.6) \textit{moderate}, $\kappa_w$ $\in$ [0.6,0.8) \textit{substantial} and $\kappa_w$ $\in$ [0.8,1] \textit{almost perfect} \cite{Landis1977}. As opposed to accuracy, $\kappa_w$ takes the rating distributions of the two sets into account, which is particularly useful when the number of ratings in each class are imbalanced. As comparison, AVRA achieved inter-rater agreements of $\kappa_w = $ 0.72 - 0.74 (left and right MTA, respectively) to an expert radiologist on a test set from the same data distribution as the training data in \cite{Martensson2018b}, similar to reported inter-rater agreements between two radiologists. Since using $\kappa_w$ required rounding AVRA's continuous predictions to the nearest integer, mean squared error (MSE) was also reported. Accuracies are included as Supplementary Data. 

\section{Results}
The rating agreements between AVRA and the neuroradiologist are summarized in Table \ref{tab:wk}. When only training on the research cohort ADNI we saw a general drop in performance in clinical cohorts compared to the test set of ADNI---particularly in the E-DLB$_{\text{C1}}$ set. Adding data from the similar cohort AddNeuroMed helped little in improving generalization, whereas the inclusion of clinical MemClin had a positive impact on performance. The overall impression was that including data from clinical cohorts in the training set improved the rating agreements and accuracies in the clinical test sets, although not consistently. Surprisingly, the rating agreement was greater in the sub-cohorts E-DLB$_{\text{DLB}}$ and E-DLB$_{\text{PDD}}$ than in E-DLB$_{\text{AD}}$ when only training on images from AD cohorts. 

\begin{table}[t!]
\caption{\label{tab:wk} Rating agreement between AVRA and a neuroradiologist expressed in Cohen's $\kappa_w$ and mean squared error (MSE) for various test sets when trained on different combinations of training cohorts. The number of training subjects were kept constant to $N=1568$ together with a fixed label distribution. A $\checkmark$ symbol in a column denotes that the cohort of that row was part of the training set. E.g. the first column shows the rating agreement and MSE for different test sets when trained only on ADNI, the second when trained on ADNI+AddNeuroMed, etc. If there was any overlap between images in a training and test set combination no agreement was computed (listed as '---' in the table). The greatest agreement values for each test set are in bold.}
\centerline{
\footnotesize
\begin{tabular}{l*{12}{c}} 
\toprule
\multicolumn{1}{c}{\underline{Cohort}} & \multicolumn{12}{c}{Cohorts incl. in training} \\ 
\cmidrule{2-13}
ADNI$^{\text{train}}$  &  \checkmark  &  \checkmark  &  \checkmark  &  \checkmark  &  \checkmark  &  \checkmark  &  \checkmark  &  \checkmark  &  \checkmark  &  \checkmark  &  \checkmark  &  \checkmark  \\ 
AddNeuroMed  &  &  \checkmark  &  &  &  &  \checkmark  &  \checkmark  &  &  &  \checkmark  &  &  \\ 
MemClin$^{\text{train}}$  &  &  &  \checkmark  &  &  \checkmark  &  &  \checkmark  &  &  &  \checkmark  &  &  \\ 
E-DLB$_{\text{C1,C2}}^{\text{train}}$  &  &  &  &  \checkmark  &  \checkmark  &  \checkmark  &  \checkmark  &  &  &  &  &  \\ 
E-DLB$_{\text{25}}^{\text{train}}$  &  &  &  &  &  &  &  &  \checkmark  &  &  &  &  \\ 
E-DLB$_{\text{50}}^{\text{train}}$  &  &  &  &  &  &  &  &  &  \checkmark  &  \checkmark  &  &  \\ 
E-DLB$_{\text{AD}}$  &  &  &  &  &  &  &  &  &  &  &  \checkmark  &  \\ 
E-DLB$_{\text{DLB}}$  &  &  &  &  &  &  &  &  &  &  &  &  \checkmark  \\ 
\addlinespace
 & & & & & \multicolumn{4}{c}{\underline{Cohen's $\kappa_w$}} \\ 
\addlinespace
ADNI$^{\text{test}}$\hfill & 0.70 & \textbf{0.71} & 0.70 & \textbf{0.71} & 0.69 & \textbf{0.71} & 0.70 & 0.69 & 0.68 & 0.70 & 0.68 & 0.69 \\ 
AddNeuroMed\hfill & \textbf{0.72} &  ---  & 0.65 & 0.65 & 0.64 &  ---  &  ---  & 0.68 & 0.66 &  ---  & 0.70 & 0.68 \\ 
MemClin\hfill & 0.65 & 0.66 &  ---  & 0.67 &  ---  & 0.68 &  ---  & 0.65 & 0.63 &  ---  & 0.69 & \textbf{0.71} \\ 
MemClin$^{\text{test}}$\hfill & 0.67 & 0.68 & 0.76 & 0.73 & 0.78 & 0.71 & 0.77 & 0.69 & 0.63 & 0.73 & 0.66 & \textbf{0.79} \\ 
E-DLB$_{\text{all}}$\hfill & 0.61 & 0.59 & \textbf{0.64} &  ---  &  ---  &  ---  &  ---  &  ---  &  ---  &  ---  &  ---  &  ---  \\ 
E-DLB$_{\text{50}}^{\text{test}}$\hfill & 0.62 & 0.59 & 0.63 &  ---  &  ---  &  ---  &  ---  & 0.63 & 0.67 & \textbf{0.68} &  ---  &  ---  \\ 
E-DLB$_{\text{AD}}$\hfill  & 0.51 & 0.54 & 0.59 &  ---  &  ---  &  ---  &  ---  &  ---  &  ---  &  ---  &  ---  & \textbf{0.65} \\ 
E-DLB$_{\text{DLB}}$\hfill & 0.63 & 0.59 & \textbf{0.65} &  ---  &  ---  &  ---  &  ---  &  ---  &  ---  &  ---  & 0.63 &  ---  \\ 
E-DLB$_{\text{PDD}}$\hfill  & \textbf{0.69} & 0.60 & 0.62 &  ---  &  ---  &  ---  &  ---  &  ---  &  ---  &  ---  & 0.65 & 0.66 \\ 
E-DLB$_{\text{C1}}$\hfill  & 0.34 & 0.33 & 0.51 & 0.38 & 0.55 & 0.55 & \textbf{0.58} &  ---  &  ---  &  ---  &  ---  &  ---  \\ 
E-DLB$_{\text{C2}}$\hfill  & 0.67 & 0.60 & \textbf{0.68} & 0.66 & \textbf{0.68} & 0.67 & 0.64 &  ---  &  ---  &  ---  &  ---  &  ---  \\ 
\addlinespace
 & & & & & \multicolumn{4}{c}{\underline{Mean squared error}} \\ 
\addlinespace
ADNI$^{\text{test}}$\hfill  & 0.26 & \textbf{0.25} & \textbf{0.25} & \textbf{0.25} & 0.26 & \textbf{0.25} & \textbf{0.25} & 0.27 & 0.27 & 0.27 & 0.26 & 0.26 \\ 
AddNeuroMed\hfill & \textbf{0.22} &  ---  & 0.24 & 0.24 & 0.25 &  ---  &  ---  & \textbf{0.22} & 0.24 &  ---  & 0.23 & 0.23 \\ 
MemClin\hfill  & 0.29 & 0.26 &  ---  & 0.25 &  ---  & \textbf{0.23} &  ---  & 0.26 & 0.26 &  ---  & 0.24 & \textbf{0.23} \\ 
MemClin$^{\text{test}}$\hfill & 0.28 & 0.23 & 0.19 & 0.21 & \textbf{0.18} & 0.21 & 0.19 & 0.24 & 0.24 & 0.21 & 0.25 & \textbf{0.18} \\ 
E-DLB$_{\text{all}}$\hfill & 0.35 & 0.35 & \textbf{0.29} &  ---  &  ---  &  ---  &  ---  &  ---  &  ---  &  ---  &  ---  &  ---  \\ 
E-DLB$_{\text{50}}^{\text{test}}$\hfill & 0.34 & 0.33 & 0.29 &  ---  &  ---  &  ---  &  ---  & 0.28 & \textbf{0.27} & \textbf{0.27} &  ---  &  ---  \\ 
E-DLB$_{\text{AD}}$\hfill  & 0.43 & 0.41 & 0.31 &  ---  &  ---  &  ---  &  ---  &  ---  &  ---  &  ---  &  ---  & \textbf{0.27} \\ 
E-DLB$_{\text{DLB}}$\hfill & 0.35 & 0.36 & \textbf{0.32} &  ---  &  ---  &  ---  &  ---  &  ---  &  ---  &  ---  & \textbf{0.32} &  ---  \\ 
E-DLB$_{\text{PDD}}$\hfill & 0.25 & 0.25 & 0.22 &  ---  &  ---  &  ---  &  ---  &  ---  &  ---  &  ---  & 0.25 & \textbf{0.21} \\ 
E-DLB$_{\text{C1}}$\hfill  & 0.75 & 0.71 & 0.40 & 0.44 & 0.38 & 0.37 & \textbf{0.36} &  ---  &  ---  &  ---  &  ---  &  ---  \\ 
E-DLB$_{\text{C2}}$\hfill  & \textbf{0.22} & 0.25 & 0.25 & 0.23 & \textbf{0.22} & 0.23 & 0.24 &  ---  &  ---  &  ---  &  ---  &  ---  \\ 
\bottomrule 
\end{tabular} 
}
\end{table}

In Fig. \ref{fig:scatter} we focus on the centers E-DLB$_\text{C1}$ and E-DLB$_\text{C2}$, where AVRA's performance metrics were particularly low (C1) or close to within-distribution test set performances (C2) when trained on research data. We compared the predictions made by the ensemble models trained only on ADNI ($x$-axis) to when trained on data from ADNI and clinical images from the MemClin and E-DLB$_\text{C1,C2}^{\text{train}}$ cohorts. Thus, no images from these centers had been part in either of the training sets, but the latter included clinical images acquired from a wider range of protocols. We observed systematic differences in the predictions between the two models, most notably in the C1 cohort. Note the intensity differences in tissue types between images from ADNI, C1 and C2 in Fig. \ref{fig:cohort_ex}. 

AVRA's MTA ratings on the test-retest cohort are plotted in Fig. \ref{fig:retest} for the models trained on the least and most heterogeneous data. We observed small intra-subject rating variability for most subjects, within the same model. It was mainly the predictions of the two images acquired with the Siemens Trio 3T that stood out. While the direction of the rating prediction differences were not consistent across subjects, it suggests that AVRA may systematically rate images acquired from some protocols/scanners differently. Comparing the two versions we see that the model trained only on ADNI systematically rates images lower than when trained also on clinical data---same as in Fig. \ref{fig:scatter}. Further, it should be noted that these participants were younger than in any of the training cohorts and---for the patients suffering from MS---from a different disease population.

\section{Discussion}
In this study we systematically showed that the performance of a CNN trained on MRI images from homogeneous research cohorts generally drops when applied to clinical data. In one center---where image intensity was visibly different to images from the training data---the performance of AVRA was lower due to a systematic underestimation. However, by including images acquired from a wider range of scanners and protocols in the training set we observed an increase in robustness/reliability of the DL model in unseen OOD data---without a substantial damage to the within-distribution test set performance. This is the first study on a large MRI neuroimaging data set labeled by the same expert neuroradiologist (thus no inter-observability bias) and with fixed training set sizes and label distribution. These results add to the evidence that rigorous testing of DL applications in medical imaging needs to be performed on external data before being used in clinics.

\begin{figure}[h!]
\centerline{
\includegraphics[height=.2\textheight]{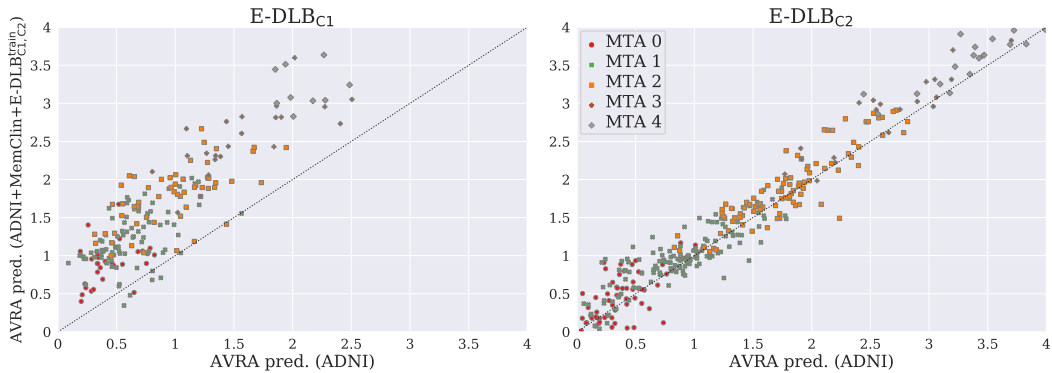}}
\caption{\label{fig:scatter} Scatter plot of the AVRA predictions of the images from E-DLB$_\text{C1}$ (left) and E-DLB$_\text{C2}$ (right), which respectively showed poor and good agreement using the baseline model trained only on ADNI. Each dot represents a subject, where the $x$-coordinate is the prediction when trained only on the ADNI cohort, and the $y$-coordinate where images from clinical cohorts were also represented in the training data. The marker symbols and colors indicate radiologist's ("ground truth") ratings. The dotted line show $x=y$, making it clear that AVRA's predictions were systematically lower than if including data from a wider distribution in the training set. This was very prominent in the E-DLB$_\text{C1}$ cohort, but also notable in E-DLB$_\text{C2}$.}
\end{figure}

From our results in Table \ref{tab:wk} we note several interesting findings. First, the level of agreement is lower in the clinical cohorts MemClin and E-DLB$_\text{all}$ when only trained on research cohorts (ADNI with or without AddNeuroMed). This suggests that we can expect a degradation of a CNN model when applied to MRI images acquired with protocols not seen during training, which is problematic for scalable deployment in clinics. Similar findings have previously been reported on segmentation tasks on cross-institutional MRI data \cite{Perone2019,Albadawy2018} and chest x-ray data \cite{Pooch2019,Yao2019}. While inter-rater agreement levels of $\kappa_w>0.6$ might be considered acceptable in many clinical situations for visual ratings (reported $\kappa_w$ levels between radiologists are typically between 0.6 and 0.8 in previous studies \cite{Martensson2018b}) we see that the agreement in E-DLB$_\text{C1}$ is substantially lower when only trained on data from harmonized research cohorts. Further, the performance drop observed in E-DLB$_\text{C1}$ but not in E-DLB$_\text{C2}$ implies that evaluating DL models in data from a single external center is not sufficient to assess the degree of generalization. In order to deploy a clinical DL model we believe that it is necessary to report the \textit{epistemic uncertainty} of a prediction, i.e. the model's uncertainty due to not having been previously exposed to a similar image during training. This would signal that more "E-DLB$_\text{C1}$-like" data needs to included in the training set for the DL model to show good performance in C1 (sometimes referred to as \textit{active learning}). Developing scalable methods to estimate DL model uncertainty---or being able to detect OOD data---is an active research field but was not explored in the current study and dataset. 

\begin{figure}[h!]
\centerline{
\includegraphics[height=.4\textheight]{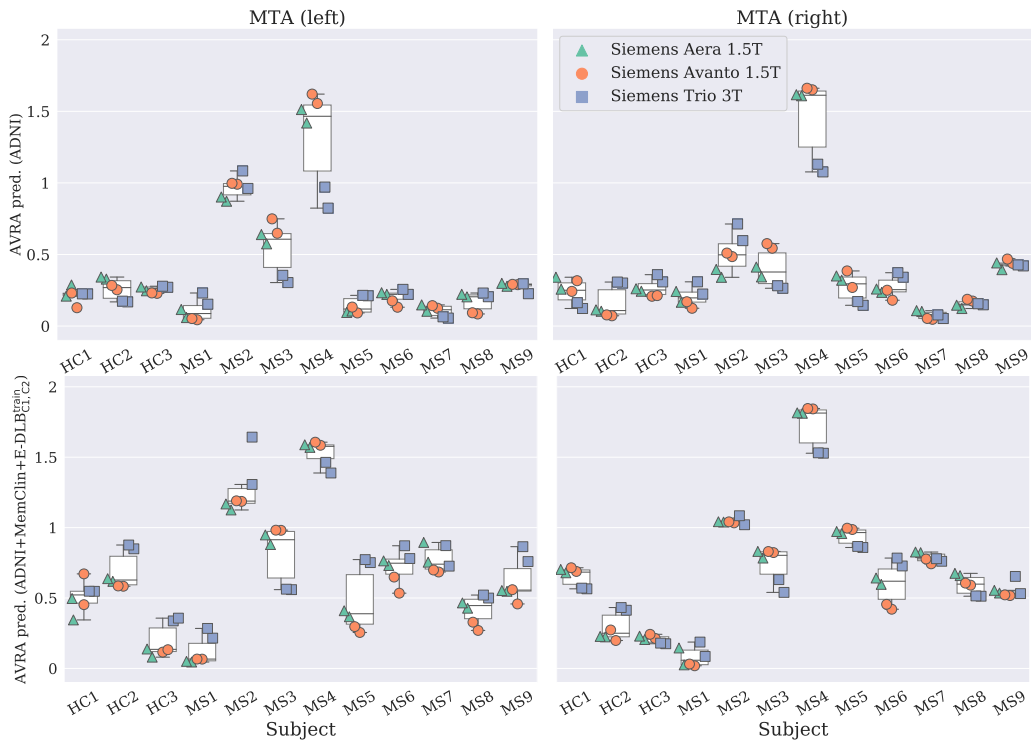}
}
\caption{\label{fig:retest} Boxplot of AVRA ratings of left MTA (left column) and right MTA (right column) for all participants in the test-retest dataset. Top row: model trained only on ADNI; Bottom row: model trained on ADNI+AddNeuroMed+MemClin+E-DLB$_{\text{C1,C2}}^{\text{train}}$. Each subject was scanned twice with repositioning in three different scanners, and each image's AVRA rating is plotted in different colors depending on scanner. Individuals denoted with the prefix HC were healthy controls and MS were patients with Multiple Sclerosis.}
\end{figure}

Second, including images of larger variability from clinical cohorts improved performance even when keeping the training set size and label distribution fixed. Including data from MemClin in the training set had a positive impact on the E-DLB sets and vice versa. This implies that by training a supervised DL model on data from a wide range of scanners, protocols, field strengths and diagnoses/labels it is possible to achieve acceptable performance on new unseen data. The systematic prediction differences for E-DLB$_{\text{C1}}$ in Fig. \ref{fig:scatter} illustrates this point well, where training data from other memory clinics had a large impact on predictions.

Third, we investigated the performance of AVRA in DLB and PDD populations when trained on images of subjects on the AD spectrum (from healthy controls, to patients with mild cognitive impairment and AD). Unexpectedly, the agreement was higher in both the DLB and PDD populations than in the AD population from the E-DLB cohort. These results could potentially be explained by the differences in rating distributions between the disease populations. PDD and DLB individuals generally had lower MTA ratings than the AD patients, and from Fig. \ref{fig:scatter} we see that the model trained only on ADNI tends to rate too low---particularly for higher MTA values. Thus, this systematic error could affect the AVRA performance in the AD population more. However, the relatively high agreements of E-DLB$_{\text{DLB}}$ and E-DLB$_{\text{PDD}}$ show potential that AVRA has the ability to generalize across disease populations. This finding is likely attributed to the strength of the clinical visual rating scales---which are disease-unspecific by design---and demonstrate the power of incorporating domain knowledge when building DL models. A previous study on applying machine learning models (SVM) on unseen clinical data reported and discussed difficulties in determining if subjects suffered from mixed pathologies (e.g. both AD and FTD) or a misdiagnosis \cite{Kloppel2015}. A model trained to discriminate between e.g. AD patients from healthy controls---both generally defined by strict inclusion and exclusion criteria in research cohorts---does just that. Applying an "AD model" like this in a more heterogeneous cohort with controls, AD and DLB subjects, would thus most probably misdiagnose DLB as AD due to resembling patterns of atrophy \cite{Oppedal2019}.

The test-retest results (Fig. \ref{fig:retest}) show impressive consistency for each DL model in most predictions. The ratings from the version trained on multiple data sets seems to yield higher variability for many subjects compared to when only trained on ADNI. Given that this model showed better generalization in the analyses summarized in Table \ref{tab:wk}, this is a bit counterintuitive. It should be noted however that these differences are small considering being trained on integer ratings with some degree of intra-rater variability. The explanation for this inter-scanner variability could partially be due to a minor overfit to scanner and protocol. This is however to prefer to the ADNI-model where the ratings seems to be systematically too low. Within-scanner and within-field strength variability was practically non-existent, and it is only the images of the 3T scanner that notably deviates for some patients. This means that we expect AVRA to be useful for longitudinal studies, where the data is typically collected in a harmonized way. Guo et al. (2019) analyzed the same dataset using different (non-machine learning) neuroimaging softwares and reported smaller within- than between-scanner variability \cite{Guo2019}. A previous study investigating the impact choice of scanner and field strength have on the performance of an SVM-classfier found the largest performance drop when training on 1.5T data and testing on 3T data and vice versa, while generalizing well to new scanners within the same field strength \cite{Abdulkadir2011}. The analyses in \cite{Abdulkadir2011} were done in the ADNI cohort, with protocols harmonized using a phantom to reduce scanner and site variability. For computer scientists it would solve many practical issues if protocols were harmonized across clinics, and that these protocols were used as default. However, this seems unlikely given the enormous effort of implementing it, the development of new (improved) sequences, and disrupting habits and workflows of clinicians. Further, the real gain of machine learning applications would be on CT images---as it is cheaper and more commonly available---where image quality variation is even greater. Thus, scanner/protocol generalization remains an important issue that needs solving prior to deploying DL models as clinical aid. Since labeled data in medicine is often difficult or expensive to acquire semi-supervised approaches may play a big role in medical machine learning applications as it allows the inclusion of unlabeled images in the training data. This has been shown to improve generalization on medical OOD data \cite{Kamnitsas2017,Perone2019,Orbes-Arteaga2019}.

The current study has some limitations that we leave as future studies. Foremost, we trained and evaluated a single network architecture and we cannot say to what degree the results are representative of DL models in general. By using the same hyperparameters as in \cite{Martensson2018b} (tuned to optimize performance on a within-distribution cross-validation set) nothing prevented AVRA from overfitting to the training protocol. Further, while the kappa metric is the most common way to quantify reliability of visual ratings, it can be noisy since we need to round the prediction to nearest integer. The MSE metric does not require rounding but is on the other hand sensitive to outliers. Since AVRA takes unprocessed MRI images as input---just as a radiologist would---we did not explore the impact of preprocessing or intensity normalization could have on generalization.

\section{Conclusion}
In this study we assessed how well a supervised deep learning model (AVRA), trained on unprocessed MRI brain images to predict Scheltens' MTA score, generalizes to external clinical data. More specifically, we trained multiple versions of AVRA on data from different combinations of research and clinical cohorts, while keeping training set size and label distribution fixed. We found that AVRA trained on homogeneous data from a research cohort generalized well to cohorts with similar protocols, but worse when applied to clinical data. On images from one specific memory clinic the performance dropped to an unacceptably low level. Including more heterogeneous data from a wider range of scanner and protocols during training improved the performance also in out-of-distribution data. Furthermore, when applying AVRA on images of patients suffering from other neurological disorders than AD we did not observe a noticeable decrease in performance. From these findings we advocate that DL models need to be rigorously tested in OOD data before being deployed in clinics. This is, to our knowledge, the largest and most comprehensive study to date on the effect of domain-shift in MRI images and deep learning models.

\clearpage

\section*{Acknowledgements}
We would like to thank the Swedish Foundation for Strategic Research (SSF), The Swedish Research Council (VR), the Strategic Research Programme in Neuroscience at Karolinska Institutet (StratNeuro), Swedish Brain Power, CIMED, Stiftelsen Olle Engkvist Byggmästare, the regional agreement on medical training and clinical research (ALF) between Stockholm County Council and Karolinska Institutet, Hjärnfonden, Alzheimerfonden, the Åke Wiberg Foundation and Birgitta och Sten Westerberg for additional financial support. The Titan X Pascal used for this research was donated by the NVIDIA Corporation.

Data collection and sharing for this project was funded by the Alzheimer's Disease Neuroimaging Initiative (ADNI) (National Institutes of Health Grant U01 AG024904) and DOD ADNI (Department of Defense award number W81XWH-12-2-0012). ADNI is funded by the National Institute on Aging, the National Institute of Biomedical Imaging and Bioengineering, and through generous contributions from the following: AbbVie, Alzheimer’s Association; Alzheimer’s Drug Discovery Foundation; Araclon Biotech; BioClinica, Inc.; Biogen; Bristol-Myers Squibb Company; CereSpir, Inc.; Cogstate; Eisai Inc.; Elan Pharmaceuticals, Inc.; Eli Lilly and Company; EuroImmun; F. Hoffmann-La Roche Ltd and its affiliated company Genentech, Inc.; Fujirebio; GE Healthcare; IXICO Ltd.; Janssen Alzheimer Immunotherapy Research \& Development, LLC.; Johnson \& Johnson Pharmaceutical Research \& Development LLC.; Lumosity; Lundbeck; Merck \& Co., Inc.; Meso Scale Diagnostics, LLC.; NeuroRx Research; Neurotrack Technologies; Novartis Pharmaceuticals Corporation; Pfizer Inc.; Piramal Imaging; Servier; Takeda Pharmaceutical Company; and Transition Therapeutics. The Canadian Institutes of Health Research is providing funds to support ADNI clinical sites in Canada. Private sector contributions are facilitated by the Foundation for the National Institutes of Health (\url{www.fnih.org}). The grantee organization is the Northern California Institute for Research and Education, and the study is coordinated by the Alzheimer’s Therapeutic Research Institute at the University of Southern California. ADNI data are disseminated by the Laboratory for Neuro Imaging at the University of Southern California.

\section*{References}
\markboth{References}{References}
\renewcommand\refname{References}
\small
\bibliography{references}{}
\newpage
\appendix
\setcounter{table}{0} \renewcommand{\thetable}{A.\arabic{table}}
\section{Supplementary data}

As supplementary data we provide accuracy results complementary to MSE and Cohen's $\kappa_w$ in Table \ref{tab:acc}. The full list of scanning parameters are found in Tables \ref{tab:adni}-\ref{tab:test_retest}.

%\ref{tab:adni,tab:addneuromed,tab:memclin,tab:edlb,tab:test_retest}
%TODO: REMOVE COHORT DESCR.?
\begin{table}[h!]
\caption{\label{tab:acc} Accuracy of (rounded) AVRA predictions and a neuroradiologist for different combinations of training cohorts and test sets. The number of training subjects were kept constant to $N=1568$ together with a fixed training distribution. A $\checkmark$ symbol in a column denotes that the cohort of that row was part of the training set. E.g. the first column shows the rating agreement and mean squared error for different test sets when trained only on ADNI, the second when trained on ADNI+AddNeuroMed, etc. If there was any overlap between images in a training and test set combination no agreement was computed (listed as '---' in the table). The greatest agreement values for each test set are in bold.}
\centerline{
\footnotesize
\begin{tabular}{l*{12}{c}} 
\toprule
\multicolumn{1}{c}{\underline{Cohort}} & \multicolumn{12}{c}{Training combination} \\ 
\cmidrule{2-13}
ADNI$^{\text{train}}$  &  \checkmark  &  \checkmark  &  \checkmark  &  \checkmark  &  \checkmark  &  \checkmark  &  \checkmark  &  \checkmark  &  \checkmark  &  \checkmark  &  \checkmark  &  \checkmark  \\ 
AddNeuroMed  &  &  \checkmark  &  &  &  &  \checkmark  &  \checkmark  &  &  &  \checkmark  &  &  \\ 
MemClin$^{\text{train}}$  &  &  &  \checkmark  &  &  \checkmark  &  &  \checkmark  &  &  &  \checkmark  &  &  \\ 
E-DLB$_{\text{C1,C2}}^{\text{train}}$  &  &  &  &  \checkmark  &  \checkmark  &  \checkmark  &  \checkmark  &  &  &  &  &  \\ 
E-DLB$_{\text{25}}^{\text{train}}$  &  &  &  &  &  &  &  &  \checkmark  &  &  &  &  \\ 
E-DLB$_{\text{50}}^{\text{train}}$  &  &  &  &  &  &  &  &  &  \checkmark  &  \checkmark  &  &  \\ 
E-DLB$_{\text{AD}}$  &  &  &  &  &  &  &  &  &  &  &  \checkmark  &  \\ 
E-DLB$_{\text{DLB}}$  &  &  &  &  &  &  &  &  &  &  &  &  \checkmark  \\ 
\addlinespace
\cmidrule{2-13}
\multicolumn{1}{c}{\underline{Accuracy}} \\ 
\addlinespace
ADNI$^{\text{test}}$\hfill & 0.68 & 0.70 & 0.68 & \textbf{0.71} & 0.68 & 0.70 & 0.69 & 0.67 & 0.68 & 0.69 & 0.66 & 0.67 \\ 
AddNeuroMed\hfill & \textbf{0.72} &  ---  & 0.65 & 0.65 & 0.64 &  ---  &  ---  & 0.68 & 0.66 &  ---  & 0.69 & 0.68 \\ 
MemClin\hfill & 0.66 & 0.67 &  ---  & 0.68 &  ---  & 0.70 &  ---  & 0.67 & 0.65 &  ---  & 0.70 & \textbf{0.72} \\ 
MemClin$^{\text{test}}$\hfill & 0.67 & 0.70 & 0.77 & 0.74 & 0.79 & 0.73 & 0.79 & 0.71 & 0.65 & 0.74 & 0.65 & \textbf{0.80} \\ 
E-DLB$_{\text{all}}$\hfill  & 0.63 & 0.61 & \textbf{0.65} &  ---  &  ---  &  ---  &  ---  &  ---  &  ---  &  ---  &  ---  &  ---  \\ 
E-DLB$_{\text{50}}^{\text{test}}$\hfill (321) & 0.64 & 0.61 & 0.65 &  ---  &  ---  &  ---  &  ---  & 0.63 & 0.68 & \textbf{0.69} &  ---  &  ---  \\ 
E-DLB$_{\text{AD}}$\hfill  & 0.56 & 0.57 & 0.60 &  ---  &  ---  &  ---  &  ---  &  ---  &  ---  &  ---  &  ---  & \textbf{0.65} \\ 
E-DLB$_{\text{DLB}}$\hfill & 0.64 & 0.61 & \textbf{0.66} &  ---  &  ---  &  ---  &  ---  &  ---  &  ---  &  ---  & 0.65 &  ---  \\ 
E-DLB$_{\text{PDD}}$\hfill & \textbf{0.75} & 0.69 & 0.69 &  ---  &  ---  &  ---  &  ---  &  ---  &  ---  &  ---  & 0.71 & 0.73 \\ 
E-DLB$_{\text{C1}}$\hfill  & 0.49 & 0.40 & 0.58 & 0.50 & 0.57 & 0.59 & \textbf{0.60} &  ---  &  ---  &  ---  &  ---  &  ---  \\ 
E-DLB$_{\text{C2}}$\hfill & 0.65 & 0.61 & \textbf{0.67} & 0.65 & \textbf{0.67} & \textbf{0.67} & 0.63 &  ---  &  ---  &  ---  &  ---  &  ---  \\ 
\bottomrule 
\multicolumn{11}{l}{\footnotesize{\textit{Cohort description:}}} \\ 
\multicolumn{11}{l}{\footnotesize{MemClin: Images from the Memory Clinic at Karolinska Hospital.}}\\ 
\multicolumn{11}{l}{\footnotesize{E-DLB$_{\text{all}}$: All images in the E-DLB cohort.}}\\ 
\multicolumn{11}{l}{\footnotesize{E-DLB$_{\text{AD,DLB}}$: Only individuals with AD (or DLB) pathology in the E-DLB cohort.}}\\ 
\multicolumn{11}{l}{\footnotesize{E-DLB$_{\text{25,50}}$: 25\% (or 50\%) of randomly sampled images from the E-DLB cohort.}}\\ 
\multicolumn{11}{l}{\footnotesize{E-DLB$_{\text{C1,C2}}^{\text{train}}$: All images from E-DLB \emph{except} from centers C1 (or C2) from E-DLB.}}\\ 
\multicolumn{11}{l}{\footnotesize{E-DLB$_{\text{C1,C2}}^{\text{test}}$: \emph{Only} images from centers C1 (or C2) from E-DLB.}}\\ 
\end{tabular} 
}
\end{table}

\newpage
\small
%\begin{longtable}{| l | c | c | c | c | c | c | c | c | c |}
\footnotesize
\begin{longtable}{l*{9}{c}}
\caption{\label{tab:adni} Detailed MRI protocols and scanners from the ADNI cohort, where each row represent an individual scanner. Abbreviations: Echo time (TE); Repetition time (TR); Inversion recovery time (IT); Slice thickness (ST); Resolution in x-y-plane (x-y).}\\
\toprule
\textbf{Scanner} & \textbf{N} & \textbf{FS} (T) & \textbf{TE} (ms) & \textbf{TR} (ms)& \textbf{IT} (ms) & \textbf{x-y} (mm)&\textbf{ST} (mm) \\
\midrule
GE Genesis signa & 114 & 1.5 & 4.10 & 10.20 & 1000 & 0.94 & 1.20\\
GE Genesis signa & 16 & 1.5 & 4.08 - 4.10 & 10.20 & 1000 & 0.94 - 1.02 & 1.20\\
GE Genesis signa & 18 & 1.5 & 4.09 & 10.40 & 1000 & 0.94 & 1.20\\
GE Genesis signa & 39 & 3 & 3.05 & 7.50 & 900 & 1.02 & 1.20\\
GE Genesis signa & 4 & 1.5 & 4.09 & 10.20 & 1000 & 0.94 & 1.20\\
GE Genesis signa & 48 & 1.5 & 4.08 - 4.10 & 10.20 & 1000 & 0.94 - 1.09 & 1.20\\
GE Genesis signa & 89 & 1.5 & 4.09 & 10.40 & 1000 & 0.94 & 1.20\\
GE Signa excite & 10 & 3.0 & 2.84 - 2.85 & 6.61 - 6.63 & 900 & 1.02 & 1.20\\
GE Signa excite & 107 & 1.5 & 3.92 - 4.05 & 8.90 - 9.20 & 1000 & 0.94 - 1.02 & 1.20\\
GE Signa excite & 14 & 1.5 & 3.80 - 3.90 & 8.59 - 8.81 & 1000 & 0.94 & 1.20\\
GE Signa excite & 155 & 1.5 & 3.80 - 3.92 & 8.59 - 8.99 & 1000 & 0.94 - 0.98 & 1.20\\
GE Signa excite & 158 & 1.5 & 3.92 - 4.05 & 8.92 - 9.20 & 1000 & 0.94 & 1.20\\
GE Signa excite & 17 & 1.5 & 3.96 & 9.12 & 1000 & 0.94 & 1.20\\
GE Signa excite & 2 & 1.5 & 3.96 & 9.12 & 1000 & 0.94 & 1.20\\
GE Signa excite & 21 & 1.5 & 3.80 & 8.59 & 1000 & 0.94 & 1.20\\
GE Signa excite & 24 & 3.0 & 2.84 - 2.85 & 6.62 - 6.63 & 900 & 1.02 & 1.20\\
GE Signa excite & 30 & 1.5 & 3.92 - 4.06 & 8.92 - 9.22 & 1000 & 0.94 & 1.20\\
GE Signa excite & 34 & 3.0 & 2.84 - 2.99 & 6.61 - 7.04 & 900 & 1.02 & 1.20\\
GE Signa excite & 39 & 1.5 & 3.80 - 3.91 & 8.60 - 8.84 & 1000 & 0.94 & 1.20\\
GE Signa excite & 45 & 1.5 & 3.96 & 9.12 & 1000 & 0.94 & 1.20\\
GE Signa excite & 46 & 1.5 & 3.80 - 3.90 & 8.59 - 8.81 & 1000 & 0.94 & 1.20\\
GE Signa excite & 51 & 1.5 & 3.92 - 4.05 & 8.92 - 9.20 & 1000 & 0.94 & 1.20\\
GE Signa excite & 52 & 1.5 & 3.92 - 3.94 & 8.92 - 8.94 & 1000 & 0.94 & 1.20\\
GE Signa excite & 56 & 1.5 & 3.78 - 3.98 & 8.56 - 9.12 & 1000 & 0.94 - 1.02 & 1.20\\
GE Signa excite & 60 & 1.5 & 3.79 - 3.91 & 8.58 - 8.84 & 1000 & 0.94 - 1.02 & 1.20\\
GE Signa excite & 8 & 3.0 & 2.86 - 3.01 & 6.64 - 6.96 & 900 & 1.02 & 1.20\\
GE Signa excite & 84 & 1.5 & 3.92 - 3.96 & 8.92 - 9.12 & 1000 & 0.94 & 1.20\\
GE Signa excite & 88 & 1.5 & 3.80 - 3.90 & 8.59 - 8.81 & 1000 & 0.94 & 1.20\\
GE Signa excite & 9 & 3.0 & 2.84 - 2.99 & 6.62 - 6.91 & 900 & 1.02 & 1.20\\
GE Signa excite & 93 & 1.5 & 3.96 & 9.12 & 1000 & 0.94 & 1.20\\
GE Signa excite & 95 & 1.5 & 3.78 - 3.91 & 8.57 - 8.84 & 1000 & 0.94 & 1.20\\
GE Signa hdx & 10 & 3.0 & 2.84 - 2.85 & 6.61 - 6.63 & 900 & 1.02 & 1.20\\
GE Signa hdx & 106 & 1.5 & 3.80 & 8.59 & 1000 & 0.94 & 1.20\\
GE Signa hdx & 149 & 1.5 & 3.79 - 4.96 & 8.58 - 11.05 & 1000 - 1044 & 0.94 - 1.09 & 1.20\\
GE Signa hdx & 16 & 1.5 & 3.80 - 3.90 & 8.59 - 8.81 & 1000 & 0.94 & 1.20\\
GE Signa hdx & 20 & 1.5 & 3.80 - 3.90 & 8.59 - 8.81 & 1000 & 0.94 & 1.20\\
GE Signa hdx & 24 & 1.5 & 3.92 & 8.92 & 1000 & 0.94 & 1.20\\
GE Signa hdx & 25 & 1.5 & 4.04 & 9.18 & 1000 & 0.94 & 1.20\\
GE Signa hdx & 50 & 1.5 & 3.80 & 8.59 & 1000 & 0.94 & 1.20\\
GE Signa hdx & 6 & 1.5 & 3.80 & 8.59 & 1000 & 0.94 & 1.20\\
GE Signa hdx & 92 & 1.5 & 3.80 - 3.92 & 8.59 - 8.99 & 1000 & 0.94 & 1.20\\
GE Signa hdx & 92 & 1.5 & 3.97 - 4.05 & 9 - 9.20 & 1000 & 0.94 & 1.20\\
GE Signa hdxt & 2 & 1.5 & 3.80 & 8.60 & 1000 & 0.94 & 1.20\\
GE Signa hdxt & 6 & 3.0 & 2.84 - 2.85 & 6.61 - 6.63 & 900 & 1.02 & 1.20\\
Philips Achieva & 12 & 3.0 & 3.05 - 3.17 & 6.39 - 6.81 & 1000 & 1.00 & 1.20\\
Philips Achieva & 24 & 3.0 & 3.25 & 6.85 & 1000 & 1.00 & 1.20\\
Philips Achieva & 35 & 3.0 & 3.13 - 3.25 & 6.78 - 6.85 & 1000 & 1.00 & 1.20\\
Philips Achieva & 65 & 1.5 & 4 - 4.01 & 8.62 & 1000 & 0.90 - 0.98 & 1.20\\
Philips Intera & 10 & 3.0 & 3.16 & 6.80 & 0 - 1000 & 1.00 & 1.20\\
Philips Intera & 17 & 3.0 & 3.16 - 3.18 & 6.80 - 6.92 & 0 - 1000 & 1.00 & 1.20\\
Philips Intera & 2 & 3.0 & 3.16 & 6.80 & 0 & 1.00 & 1.20\\
Philips Intera & 26 & 3.0 & 3.13 - 3.25 & 6.76 - 6.84 & 1000 & 1.00 & 1.20\\
Philips Intera & 34 & 1.5 & 3.99 - 4.01 & 8.58 - 8.62 & 0 - 1000 & 0.94 - 0.98 & 1.20\\
Philips Intera & 38 & 1.5 & 4 & 8.62 & 0 - 1000 & 0.94 & 1.20\\
Philips Intera & 4 & 1.5 & 4 & 8.62 & 1000 & 0.94 & 1.20\\
Philips Intera & 4 & 3.0 & 3.16 & 6.80 - 6.81 & 1000 & 1.00 & 1.20\\
Philips Intera & 40 & 1.5 & 3.98 - 4.01 & 8.51 - 8.62 & 1000 & 0.94 & 1.20\\
Philips Intera & 48 & 3.0 & 3.16 & 6.80 - 6.81 & 1000 & 1.00 & 1.20\\
Philips Intera & 54 & 1.5 & 3.99 - 4.01 & 8.51 - 8.63 & 1000 & 0.94 - 0.98 & 1.20\\
Philips Intera & 6 & 1.5 & 4 - 4.01 & 8.61 & 1000 & 0.94 & 1.20\\
Philips Intera & 6 & 3.0 & 3.13 - 3.25 & 6.76 - 6.84 & 1000 & 1.00 & 1.20\\
Philips Intera & 61 & 1.5 & 4 & 8.61 - 8.62 & 1000 & 0.94 & 1.20\\
Philips Intera & 66 & 1.5 & 4 & 8.59 - 8.61 & 1000 & 0.94 & 1.20\\
Philips Intera achieva & 4 & 1.5 & 3.98 & 8.55 & 1000 & 0.94 & 1.20\\
Siemens Allegra & 12 & 2.9 & 2.91 & 2300 & 900 & 1.00 & 1.20\\
Siemens Allegra & 54 & 2.9 & 2.91 & 2300 & 900 & 1.00 & 1.20\\
Siemens Allegra & 72 & 2.9 & 2.91 & 2300 & 900 & 1.00 & 1.20\\
Siemens Avanto & 109 & 1.5 & 3.50 & 2400 & 1000 & 1.25 & 1.20\\
Siemens Avanto & 30 & 1.5 & 3.50 & 2400 & 1000 & 1.25 & 1.20\\
Siemens Avanto & 30 & 1.5 & 3.52 - 3.54 & 2400 & 1000 & 1.25 - 1.30 & 1.20\\
Siemens Avanto & 4 & 1.5 & 3.50 & 2400 & 1000 & 1.25 & 1.20\\
Siemens Avanto & 44 & 1.5 & 3.50 & 2400 & 1000 & 1.25 & 1.20\\
Siemens Avanto & 52 & 1.5 & 3.52 - 3.54 & 2400 & 1000 & 1.25 - 1.30 & 1.20\\
Siemens Avanto & 63 & 1.5 & 3.50 & 2400 & 1000 & 1.25 & 1.20\\
Siemens Espree & 8 & 1.5 & 3.59 & 2400 & 1000 & 1.25 & 1.20\\
Siemens Sonata & 116 & 1.5 & 3.54 & 2400 & 1000 & 1.25 & 1.20\\
Siemens Sonata & 117 & 1.5 & 3.54 & 2400 & 1000 & 1.25 & 1.20\\
Siemens Sonata & 63 & 1.5 & 3.54 & 3000 & 1000 & 1.25 & 1.20\\
Siemens Sonata & 72 & 1.5 & 3.54 & 3000 & 1000 & 1.25 & 1.20\\
Siemens Sonata & 77 & 1.5 & 3.54 - 3.55 & 3000 & 1000 & 1.25 & 1.20\\
Siemens Sonatavision & 2 & 1.5 & 3.57 & 3000 & 1000 & 1.25 & 1.20\\
Siemens Sonatavision & 30 & 1.5 & 3.54 & 2400 & 1000 & 1.25 & 1.20\\
Siemens Symphony & 100 & 1.5 & 3.61 & 3000 & 1000 & 1.25 & 1.20\\
Siemens Symphony & 104 & 1.5 & 3.61 - 3.67 & 3000 & 1000 & 1.25 & 1.20\\
Siemens Symphony & 2 & 1.5 & 3.61 & 3000 & 1000 & 1.25 & 1.20\\
Siemens Symphony & 20 & 1.5 & 3.61 & 3000 & 1000 & 1.25 & 1.20\\
Siemens Symphony & 29 & 1.5 & 3.61 & 3000 & 1000 & 1.25 & 1.20\\
Siemens Symphony & 30 & 1.5 & 3.67 - 3.71 & 3000 & 1000 & 1.25 - 1.35 & 1.20\\
Siemens Symphony & 58 & 1.5 & 3.61 - 3.79 & 3000 & 1000 & 1.25 - 1.35 & 1.20\\
Siemens Symphony & 81 & 1.5 & 2.88 - 3.65 & 3000 & 1000 & 1.25 & 1.20\\
Siemens Symphony & 92 & 1.5 & 3.87 & 3000 & 1000 & 1.25 & 1.20\\
Siemens Symphony & 97 & 1.5 & 3.59 - 3.63 & 3000 & 1000 & 1.25 - 1.30 & 1.20\\
Siemens Symphonytim & 110 & 1.5 & 3.64 & 3000 & 1000 & 1.25 & 1.20\\
Siemens Trio & 140 & 3.0 & 2.94 & 2300 & 900 & 1.00 & 1.20\\
Siemens Trio & 4 & 2.9 & 2.94 & 2300 & 900 & 1.00 & 1.20\\
Siemens Trio & 44 & 2.9 & 2.94 & 2300 & 900 & 1.00 & 1.20\\
Siemens Trio & 5 & 3.0 & 2.94 & 2300 & 900 & 1.00 & 1.20\\
Siemens Trio & 55 & 2.9 & 2.94 & 2300 & 900 & 1.00 & 1.20\\
Siemens Trio & 6 & 2.9 & 2.94 & 2300 & 900 & 1.00 & 1.20\\
Siemens Trio & 6 & 2.9 & 2.94 & 2300 & 900 & 1.00 & 1.20\\
Siemens Trio & 6 & 3.0 & 2.96 & 2300 & 900 & 1.00 & 1.20\\
Siemens Trio & 9 & 3.0 & 2.94 & 2300 & 900 & 1.00 & 1.20\\
Siemens Triotim & 12 & 3.0 & 2.86 - 2.91 & 2300 & 900 & 1.00 - 1.03 & 1.20\\
Siemens Triotim & 12 & 3.0 & 2.91 & 2300 & 900 & 1.00 & 1.20\\
Siemens Triotim & 13 & 3.0 & 2.91 & 2300 & 900 & 1.00 & 1.20\\
Siemens Triotim & 15 & 3.0 & 2.86 - 2.91 & 2300 & 900 & 1.00 & 1.20\\
Siemens Triotim & 16 & 3.0 & 2.91 & 2300 & 900 & 1.00 & 1.20\\
Siemens Triotim & 2 & 3.0 & 2.91 & 2300 & 900 & 1.00 & 1.20\\
Siemens Triotim & 2 & 3.0 & 2.91 & 2300 & 900 & 1.00 & 1.20\\
Siemens Triotim & 2 & 3.0 & 2.91 & 2300 & 900 & 1.00 & 1.20\\
Siemens Triotim & 22 & 3.0 & 2.86 - 2.91 & 2300 & 900 & 1.00 & 1.20\\
Siemens Triotim & 4 & 3.0 & 2.91 & 2300 & 900 & 1.00 & 1.20\\
\midrule
\end{longtable}

\begin{table}[h!]
\caption{\label{tab:addneuromed} Scanners and scanning parameters of the AddNeuroMed cohort. Abbreviations: Echo time (TE); Repetition time (TR); Inversion recovery time (IT); Slice thickness (ST); Resolution in x-y-plane (x-y, equal lengths). }
\centerline{
\footnotesize
\begin{tabular}{l c c c c c c c}
\toprule
\textbf{Scanner} & \textbf{N} & \textbf{FS} (T) & \textbf{TE} (ms) & \textbf{TR} (ms)& \textbf{IT} (ms) & \textbf{x-y} (mm)&\textbf{ST} (mm) \\
\midrule
GE Genesis Signa &7 & 1.5  & 4.09  & 10.20& 1000 & 0.94 & 1.20 \\ 
GE Signa HDx &4 & 1.5  & 4.10& 10.20 & 1000 & 0.94 & 1.20 \\ 
GE Genesis Signa &24 & 1.5  & 4.09& 10.40 & 1000 & 0.94 & 1.20 \\ 
GE Genesis Signa &31 & 1.5  & 4.09& 10.20 & 1000 & 0.94 & 1.20 \\ 
GE Signa HDx &10 & 1.5  & 3.80& 8.59-8.60 & 1000 & 0.94-1.02 & 1.20 \\ 
Siemens Avanto &28 & 1.5  & 3.50 & 2400& 1000 & 1.17 & 1.20 \\ 
Picker Edge 1.5T &18 & 1.5 & 3.00 & 13.00 & --- & 0.94 & 1.20 \\
\bottomrule
\end{tabular}}
\end{table}

\begin{table}[h!]
\caption{\label{tab:memclin} Detailed MRI protocols and scanners from the MemClin cohort, where each row represent an individual scanner. Abbreviations: Echo time (TE); Repetition time (TR); Inversion recovery time (IT); Slice thickness (ST); Resolution in x-y-plane (x-y, equal lengths). }
\centerline{
\footnotesize
\begin{tabular}{l c c c c c c c}
%\centerline{\small
\toprule
\textbf{Scanner} & \textbf{N} & \textbf{FS} (T) & \textbf{TE} (ms) & \textbf{TR} (ms)& \textbf{IT} (ms) & \textbf{x-y} (mm)&\textbf{ST} (mm) \\
\midrule
Siemens Aera & 2 & 1.5 & 3.18 & 2300 & 904 & 1.25 & 1.20\\
Siemens Avanto & 204 & 1.5 & 2.56 - 4.19 & 1160 - 2400 & 600 - 1100 & 0.45 - 1.30 & 1.20 - 2.50\\
Siemens Magnetom vision & 36 & 1.5 & 4.40 & 11.40 & 300 & 0.90 & 2.50 - 2.50\\
Siemens Symphony & 85 & 1.5 & 3.93 & 1960 & 790 - 875 & 0.49 & 1.41 - 1.51\\
Siemens Triotim & 57 & 3.0 & 2.57 - 3.42 & 1780 - 2300 & 900 & 0.85 - 1 & 1 - 1.40\\
\bottomrule
\end{tabular}}
\end{table}

\begin{table}[h!]
\caption{\label{tab:edlb} Scanners and scanning parameters of the E-DLB cohort. The linespaces in the table are used to separate between clinics. Abbreviations: Echo time (TE); Repetition time (TR); Inversion recovery time (IT); Slice thickness (ST); Resolution in x-y-plane (x-y, equal lengths). }
\centerline{
\footnotesize
\begin{tabular}{l c c c c c c c}
\toprule
 \textbf{Scanner} & \textbf{N} & \textbf{FS (T)} & \textbf{TR (ms)} & \textbf{TE (ms)} & \textbf{IT (ms)} & \textbf{Res. (mm)} & \textbf{ST (mm)} \\  
\midrule
Philips Intera Achieva (E-DLB$_{\text{C1}}$) & 101 & 3 & 8.3 & 4.6 & 1250 & 1.00 & 1.00 \\ 
\addlinespace
Siemens Verio (E-DLB$_{\text{C2}}$) & 165 & 3 & 1900 & 2.53 & 900 & 1.00 & 1.00 \\ 
\addlinespace
BRE Siemens Avanto &24 & 1.50 & 2050 & 2.56 & 1100 & 0.50 & 1.00 \\ 
\addlinespace
GE Discovery MR750 &3 & 3 & 8.16-9.82 & 3.18-4.31 & 450 & 0.47-1.00 & 1.00-1.40 \\ 
Siemens Symphony & 9 & 1.50 & 1630-1700 & 3.93 & 1100 & 0.48-0.59 & 1.00-1.17 \\ 
\addlinespace
Philips Achieva & 9 & 3 & 10.75 & 5.07 & --- & 0.80 & 0.80 \\ 
Philips --- & 15 & --- & 19.00 & 3.68-3.75 & --- & 0.80 & 1.00 \\ 
\addlinespace
GE Signa Excite & 6 & 1.5 & 22.00 & 7.00 & --- & 1.02-1.09 & 1.60 \\ 
GE Signa HDxt & 11 & 1.5 & 11.14-11.44 & 4.98-5.15 & --- & 0.50 & 1.00 \\ 
Philips Achieva & 10 & 1.5 & 25 & 5.28-5.42 & --- & 0.45 & 0.80 \\ 
\addlinespace
Philips Achieva & 17 & 1.5 & 7.07-25 & 3.21-4.61 & --- & 0.94-1.00 & 1.00 \\ 
Siemens TrioTim & 5 & 3 & 2300 & 4.68-4.71 & 1100 & 1.00 & 1.00-1.09 \\ 
\addlinespace
GE Signa Excite & 41 & 1.5 & 8.22 & 3.12 & 500-1981 & 0.50-1.10 & 1.00-1.10 \\ %??? IT correct?
Philips Intera & 1 & 1.5 & 20 & 4.60 & --- & 0.98 & 1.00 \\ 
Philips Intera & 99 & 1.5 & 7.10 & 3.21 & --- & 1.00 & 1.00 \\ 
\addlinespace
Siemens Avanto & 59 & 1.5 & 1820 & 3.04 & 1100-2500 & 0.96-1.00 & 1.00-1.04 \\ 
\addlinespace
Philips Achieva &13 & 1.5 & 8.54 & 3.988 & --- & 0.94 & 1.20 \\ 
\addlinespace
Philips Achieva &12 & 3 & 13.03-13.15 & 7.31-7.35 & --- & 0.94 & 1.00 \\ 
\addlinespace
--- & 45 & 1.5 & --- & --- & --- & 0.60-0.87 & 0.90-1.04 \\ 
\bottomrule
\end{tabular}
}
\end{table}

\begin{table}[h!]
\caption{\label{tab:test_retest} Scanner and scanning parameter of the test-retest cohort. Abbreviations: Echo time (TE); Repetition time (TR); Inversion recovery time (IT); Phase angle (PA); Slice thickness (ST); Resolution in x-y-plane (x-y, equal lengths). }
\centerline{
\footnotesize
\begin{tabular}{l c c c c c c c c}
\toprule
 \textbf{Scanner} & \textbf{N} & \textbf{FS (T)} & \textbf{TR (ms)} & \textbf{TE (ms)} & \textbf{IT (ms)} & \textbf{FA ($^\circ$)} & \textbf{Res. (mm)} & \textbf{ST (mm)} \\  
\midrule
Siemens Aera & 24 & 1.5 & 1900 & 3.02 & 1100 & 15 & 1.0 & 1.5  \\ 
Siemens Avanto & 24 & 1.5 & 1900 & 3.55 & 1100 & 15 & 1.0 & 1.5  \\ 
Siemens Trio & 24 & 3 & 1900 & 3.39 & 900 & 9 & 1.0 & 1.5  \\ 
\bottomrule
\end{tabular}
}
\end{table}

\end{document}